\documentclass[aps,pra,showpacs,showkeys,twocolumn,groupedaddress]{revtex4-1}

\usepackage{graphicx,tabularx,times,hyperref}

\setcitestyle{numbers,square}
\usepackage{array}
\usepackage{amsmath}
\usepackage{units}
\usepackage{graphicx}
\usepackage{braket}
\usepackage{xfrac}
\usepackage{enumitem}
\usepackage[normalem]{ulem}
\usepackage{booktabs}
\usepackage{tabularx}
\usepackage{multirow, bigstrut}
\usepackage{hhline}
\usepackage{color,soul}
\usepackage{times}
\usepackage{hyperref}

\hypersetup{
colorlinks=true,
linkcolor=cyan,
filecolor=magenta, 
urlcolor=blue,
}
\begin{document}

%\title{Title}
%\author{First Author}
%\email{Author email}
%\affiliation{First Author Affiliation}
%\author{Second Author}
%\author{Third Author}
%\affiliation{Other Affiliation}

\title{Student difficulties with the corrections to the energy spectrum of the hydrogen atom for the intermediate field Zeeman effect}
%involving the Zeeman effect in the hydrogen atom}
%\pacs {01.40Fk,01.40.gb,01.40G-,1.30.Rr}
% \keywords {a}
\author{Emily Marshman}
\author{Christof Keebaugh}
\author{Chandralekha Singh}
\affiliation{Department of Physics and Astronomy, University of Pittsburgh, Pittsburgh, PA 15260}
%\date{\today}

\begin{abstract}
We discuss an investigation of student difficulties with the corrections to the energy spectrum of the hydrogen atom for the intermediate field Zeeman effect using the degenerate perturbation theory. The investigation was carried out in advanced quantum mechanics courses by administering free-response and multiple-choice questions and conducting individual interviews with students. We find that students share many common difficulties related to relevant physics concepts. In particular, students often struggled  with mathematical sense-making in this context of quantum mechanics which requires interpretation of the implications of degeneracy in the unperturbed energy spectrum and how the Zeeman perturbation will impact the splitting of the energy levels. We discuss how  the common difficulties often arise from the fact that applying linear algebra concepts correctly in this context with degeneracy in the energy spectrum is challenging for students.% of quantum mechanics.   %We describe how the research on student difficulties was used as a guide to develop and evaluate a Quantum Interactive Learning Tutorial (QuILT), which strives to help students develop a functional understanding of concepts necessary for finding the corrections to the energy spectrum of the hydrogen atom for the intermediate field Zeeman effect using the DPT. We also discuss the development of the DPT QuILT focusing on these issues and its evaluation in the undergraduate and graduate courses. 
\end{abstract}

\maketitle

\section{Introduction and Background}
%\vspace*{-.15in}
%It is important to help students develop a functional understanding of DPT in order to find the corrections to the energies of the hydrogen atom for the Zeeman effect. However, 
Quantum mechanics (QM) is challenging even for upper-level undergraduate and graduate students and students often struggle with the non-intuitive subject matter and in making connections between mathematics and QM concepts (e.g., see Refs. \cite{singh, zollman,  singh4,singh45,singh2,singh37,marshman2, singh5,wittmann, singh3}).  %\cite{singh, kohnle, kohnle2, muller, domert, singh2,zollman, maries, marshman3, siddiqui,zhu2,zhu7,zhu9,lin, singh4}). % \cite{singh,belloni,wittmann,zollman,singh4,singh5} 
%There have been a number of prior research studies aimed at investigating student reasoning in QM \cite{wittmann,marshman2,marshman4, marshman5,zhu6,singh3,singh5, singh7, brown2, marshman6, mason} and using the findings as resources for improving student understanding \cite{singh44, zhu, zhu3,zhu4,zhu5,zhu62,singh6, singh8, brown}. 
%Guided by research studies conducted to identify student difficulties with QM and findings of cognitive research, we have been developing a set of research-based learning tools including the Quantum Interactive Learning Tutorials (QuILTs) which strive to help students develop a solid grasp of QM \cite{singh33,devore,marshman,marshman7,marshman8,zhu10,brown3,marshman9,marshman10,marshman11,sayer}.  
%However, there has been relatively little research that focuses on student understanding of advanced topics in quantum mechanics, e.g., degenerate perturbation theory (DPT) \cite{keebaugh}. Here we discuss an investigation of student difficulties with mathematical sense-making in a physical situation in the context of DPT involving the Zeeman effect for the hydrogen atom. %We also describe the development and validation of the research-based QuILT that uses student difficulties as resources and strives to help students learn to apply linear algebra mathematical concepts correctly to find the corrections to the energy spectrum of the hydrogen atom for the Zeeman effect.
%Prior research suggests that students often have difficulty applying mathematical concepts in the context of a concrete physical problem. 
Prior research studies have found that students have difficulty connecting and applying mathematics correctly in introductory physics contexts (e.g., see Refs. \cite{maries4, li,tuminaro}).  Mathematical sense-making in the context of solving physics problems can often be more difficult than when solving equivalent mathematics problems without the physics context \cite{maries4, li, tuminaro}.  Since working memory is constrained to a limited number of chunks and students' knowledge chunks pertaining to physics concepts are small when they are developing expertise, use of mathematics in physics can increase the cognitive load during problem solving especially if students are not proficient in the mathematics involved \cite{sweller}.  Therefore, students may struggle to integrate mathematics and physics concepts.  Since mathematical sense-making while focusing on solving a physics problem is often more challenging, students sometimes make mathematical mistakes that they otherwise would not make if the physics context was absent \cite{maries4, li, tuminaro}.
 
One QM concept that involves connecting mathematics to a physical situation is degenerate perturbation theory (DPT) in the context of the Zeeman effect.  We investigated student difficulties with finding the first-order corrections to the energies of the hydrogen atom for the Zeeman effect using DPT, which included probing of difficulties in mathematical sense making in this QM context so that the research can be used as a guide to develop learning tools to improve student understanding.  
While the solution for the Time-Independent Schr\"{o}dinger Equation (TISE) for the hydrogen atom with Coulomb potential energy can be obtained exactly, the TISE involving the Zeeman effect must also include the fine structure correction and cannot be solved exactly. %The solution for the TISE for the hydrogen atom with Coulomb potential energy   %, predicted by the simplified Bohr model. 
%However, the TISE for the Hamiltonian with the fine structure and Zeeman corrections cannot be solved exactly. Nevertheless, 
%Since there is degeneracy in the energy spectrum of the hydrogen atom, DPT must be used.
%Perturbation theory is a particularly powerful approximation method for systems that are not exactly solvable. 
%To account for fine structure corrections to the energy spectrum of hydrogen, one must use perturbation theory. 
However, since the fine-structure and, in general, the Zeeman corrections to the energies are significantly smaller than the unperturbed energies, perturbation theory is an excellent method for computing the corrections to the energies.  % and comparing the theoretical results with experiments. %It is often the case that the symmetry of the situation creates degeneracy in the energy spectrum and this degeneracy must be accounted for in order to describe the energy shifts due to the interactions correctly. 
The high degree of symmetry of the unperturbed Hamiltonian leads to degeneracy in the energy spectrum and DPT must be used to find the perturbative corrections for the Zeeman effect. 
%In the context of degenerate perturbation theory (DPT) when the unperturbed Hamiltonian and the perturbation $\hat{H}'$ operators commute, the degeneracy in the energy spectrum of the unperturbed Hamiltonian $\hat{H}^0$ requires that students have a deep understanding of (1) bases that make both the unperturbed Hamiltonian and the perturbation $\hat{H}'$ matrices diagonal and (2) how to find a basis that diagonalizes $\hat{H}^0$ and diagonalizes $\hat{H}'$ in the degenerate subspace of $\hat{H}^0$ if a complete set of simultaneous eigenstates cannot be easily identified in order to apply it appropriately to solve problems involving the hydrogen atom.

%\vspace*{-.25in}
%\section{Background} 
%\vspace*{-.15in}
%Below we discuss the basics of DPT that %and what students should be able to do after working through 
%the QuILT strives to help students learn in the context of the Zeeman effect. In particular, students should be able to identify a {\it good} basis for finding the perturbative corrections to the energies for the Zeeman effect, which includes corrections due to both the fine structure and Zeeman terms.

%PT is a useful approximation method for finding the energies and the energy eigenstates for a system for which the TISE is not exactly solvable. 
The Hamiltonian $\hat{H}$ for the system can be expressed as the sum of two terms, the unperturbed Hamiltonian $\hat{H}^0$ and the perturbation $\hat{H}'$, i.e., $\hat{H}=\hat{H}^0+\hat{H}'$. The TISE for the unperturbed Hamiltonian, $\hat{H}^0\psi_n^0 = E_n^0\psi_n^0$, is exactly solvable, where $\psi_n^0$ is the $n^{th}$ unperturbed energy eigenstate and $E_n^0$ is the unperturbed energy. %PT builds on the solutions of the TISE for the unperturbed case. 
The $n^{th}$ energy can be approximated as $E_n =E_n^0 + E_n^1+E_n^2 + \ldots$ where $E_n^i$ for $i=1,2,3..$ are the $i^{\textrm th}$ order corrections to the $n^{\textrm th}$ energy of the system. 
%The energy eigenstate can be approximated as $\psi_n = \psi_n^0+\psi_n^1+\psi_n^2 + \cdots$ where $\psi_n^i$ are the 
% $i^{\textrm th}$ order corrections to the n$^{\textrm th}$ energy eigenstate. 
%We focus on the first-order perturbative corrections to the energy since they are usually the dominant corrections. 
In perturbation theory, the first-order corrections to the energies are
%\vspace*{-.01in}
%\begin{equation}\label{energy}
$E_n^1 = \langle \psi_n^0|\hat{H}'|\psi_n^0\rangle$
%\end{equation} 
%\vspace*{-.01in}
and the first-order corrections to the unperturbed energy eigenstates are 
%\vspace*{-.05in}
%\begin{equation}\label{wave}
$|\psi_n^1\rangle = \sum_{m \neq n} \frac{\langle \psi_m^0|\hat{H}'|\psi_n^0\rangle }{(E_n^0-E_m^0)}|\psi_m^0\rangle$, in which
%\end{equation}
%\vspace*{-.01in}
%In Eqs. \ref{energy} and \ref{wave}, 
$\left\{|\psi_n^0\rangle \right\}$ is a complete set of eigenstates of the unperturbed Hamiltonian 
$\hat{H}^0$. 
If the eigenvalue spectrum of $\hat{H}^0$ has degeneracy, %two or more eigenstates of $\hat{H}^0$ have the same energy, 
%i.e., two or more diagonal elements of $\hat{H}^0$ are equal), %Eq. \ref{energy} from 
the perturbative corrections are only valid provided one uses a {\it good} basis \cite{griffiths}. 
 For a given $\hat{H}^0$ and $\hat{H}'$, a {\it good} basis consists of a complete set of eigenstates of $\hat{H}^0$ that diagonalizes $\hat{H}'$ in each degenerate subspace of $\hat{H}^0$.
% (while keeping $\hat{H}^0$ diagonal). 
%In a {\it good} basis, $\hat{H}'$ is diagonal in each degenerate subspace of $\hat{H}^0$. 
%Therefore, the terms $\langle \psi_m^0|\hat{H}'|\psi_n^0\rangle$ in Eq. \ref{energy} for the wavefunction are zero when $m \neq n$ so that the expression for the corrections to the wavefunction in Eq. \ref{energy} do not have terms that diverge. In a {\it good} basis, Eq. \ref{energy} is also valid for finding the first-order corrections to the energies (which are the diagonal elements of the $\hat{H}'$ matrix as given by Eq. \ref{energy}).
%\subsection{Background for DPT involving the Zeeman effect}
%\vspace*{-.15in}

%For a hydrogen atom in an external magnetic field, one can use DPT 
%that students often learn about in upper-level undergraduate and graduate QM courses 
%to find the corrections to the energy spectrum. 
%As an application of DPT, here we focus on the example of a hydrogen atom placed in an external magnetic field, a topic that one part of the DPT QuILT focused on. 
%Using standard notation with $\epsilon \ll 1$, 
Using standard notations, the unperturbed Hamiltonian $\hat{H}^0$ of a hydrogen atom %placed in an external magnetic field is 
%\begin{equation}
%$\hat{H}= \hat{H}^0 + \epsilon\hat{H}'$
% = \hat{H}^0 + \epsilon(\hat{H}'_{r} + \hat{H}'_{SO} +\hat{H}'_{Z})$ %=\hat{H}^0 + \hat{H}'_{fs} +\hat{H}'_{Z}$
%\end{equation}
%in which the unperturbed Hamiltonian, 
is $\hat{H}^0 = \frac{\hat{p}^2}{2m}-\frac{e^2}{4 \pi \epsilon_0}\frac{1}{r}$, which accounts only for the interaction of the electron with the nucleus via Coulomb attraction. The solution for the TISE for the hydrogen atom with Coulomb potential energy gives the unperturbed energies $E_n^0=-\frac{13.6\textrm{eV}}{n^2}$, where $n$ is the principal quantum number.  %$E_n^0=-\frac{13.6\textrm{eV}}{n^2}$, where $n$ is the principal quantum number. %The perturbation is $\hat{H}' =\hat{H}'_{fs}+\hat{H}'_Z$, in which $\hat{H}'_{Z}$ is the Zeeman term and $\hat{H}'_{fs}$ is the fine structure term.  
The perturbation is $\hat{H}' =\hat{H}'_{fs}+\hat{H}'_Z$, in which $\hat{H}'_{Z}$ is the Zeeman term and $\hat{H}'_{fs}$ is the fine structure term.  %The Zeeman term accounts for the potential energy of the magnetic moments due to the orbital and spin angular momenta in the external magnetic field.  
The Zeeman term is given by $\hat{H}'_{Z}=\frac{\mu_B B_{ext}}{\hbar}(\hat{L}_z+2\hat{S}_z)$ in which $\vec{B}_{ext}= B_{ext}\hat{z}$ is a uniform, time independent external magnetic field along the $\hat{z}$-direction, $\mu_B$ is the Bohr magneton and $\hat{L}_z$ and $\hat{S}_z$ are the operators corresponding to the z component of the orbital and spin angular momenta, respectively.  The fine structure term includes the spin-orbit coupling  and a relativistic correction for the kinetic energy, and is expressed as $\hat{H}'_{fs} = \hat{H}'_r+\hat{H}'_{SO}$.   Here, $\hat{H}'_r=-\frac{\hat{p}^4}{8m^3c^2}$ is the relativistic correction term and $\hat{H}'_{SO}=\frac{e^2}{8 \pi \epsilon_0}\frac{1}{m^2 c^2 r^3}\vec{S}\cdot \vec{L}$ is the spin-orbit interaction term (all notations are standard).
We note that the unperturbed Hamiltonian is spherically symmetric since $[\hat{H}^0,\hat{\vec{L}}]=0$.  Therefore, for a fixed $n$, $\hat{H}^0$ for the hydrogen atom is %spherically symmetric and is 
diagonal when any complete set of orthogonal states is chosen for the angular part of the basis (consisting of the product states of orbital and spin angular momenta). Thus, so long as the radial part of the basis is always chosen to be a stationary state wavefunction $R_{nl}(r)$ for the unperturbed hydrogen atom (for given principal and azimuthal quantum numbers $n$ and $l$),  which we will assume throughout, %the radial part of the wavefunctions. %corresponding to the eigenstates of $\hat{H}^0$ is chosen as the basis, 
the choice of a {\it good} basis amounts to choosing the angular part of the basis (angular basis) %(the part of the basis that involves the product states of the orbital and spin angular momenta) 
appropriately, i.e., ensuring that the perturbation is diagonal in each degenerate subspace of $\hat{H}^0$. Therefore, we focus on the angular basis to find a {\it good} basis and the corrections to the energies for the Zeeman effect.  %Therefore, for the operator $\hat{H}^0$ in question Q2, options i, ii, iii, and iv are all correct. 
%In DPT, a {\it good} basis must diagonalize $\hat{H}^0$ and simultaneously diagonalize the perturbation $\hat{H}'$ in each degenerate subspace of $\hat{H}^0$. 
%For a given value of $n$, the $2n^2$-dimensional subspace of $\hat{H}^0$ is $2n^2$-fold degenerate and therefore the perturbation must also be diagonal for each subspace corresponding to the given $n$. 
%Since a {\it good} basis must diagonalize $\hat{H}^0$ and $\hat{H}'$ in the subspace corresponding to each value of $n$, students must determine a basis that
%which of the options i, ii, iii, or iv, if any, 
%diagonalizes $\hat{H}'$ for the given $n$.
For the angular basis for each $n$, states in the coupled representation $|l,j, \ m_j \rangle$ are labeled by the quantum numbers  $ l,  \ j$, and $m_j$ (they are eigenstates of $\hat J^2$ and $\hat J_z$) and the total angular momentum is defined as $\vec{J}=\vec{L}+\vec{S}$ (all notations are standard and $s$ has been suppressed from the states $|l,j, \ m_j \rangle$ since $s=1/2$ is a fixed value for a hydrogen atom).  %and the eigenvalue equations corresponding to the total angular momentum operator for states in the coupled representation are $\hat{J}^2|j \ m_j\rangle = \hbar^2 j(j+1)|j \ m_j\rangle$ and $\hat{J}_z|j \ m_j\rangle = \hbar m_j |j \ m_j\rangle$. %Additionally, the following eigenvalue equations are important for the strong and weak field Zeeman effect when basis states are chosen in the coupled representation: $\hat{S}^2|j \ m_j\rangle = \hbar^2 s(s+1)|j \ m_j\rangle$ and $\hat{L}^2|j \ m_j\rangle = \hbar^2 l(l+1)|j \ m_j\rangle$. 
On the other hand, states {\small $|l,\ m_l,\ m_s \rangle$} in the uncoupled representation are labeled by the quantum numbers  $ l, \ m_l,$ and $m_s$  (notations are standard) and are eigenstates of $\hat L_z$ and $\hat S_z$.

An angular basis consisting of states in the coupled representation forms a {\it good} basis for the fine structure term $\hat{H}'_{fs}$ since with this choice of the angular basis, $\hat{H}'_{fs}$ is diagonal in each degenerate subspace of $\hat{H}^0$.  But a basis consisting of states in the uncoupled representation forms a {\it good} basis for the Zeeman perturbation $\hat{H}'_{Z}$. (In this case with $\hat{H}'_Z$ only, first order PT yields the exact result since $[\hat{H}^0, \hat{H}'_Z]=0$.) Therefore, for the intermediate field Zeeman effect, in which $\hat{H}'=\hat{H}'_{fs} + \hat{H}'_{Z}$ and $\hat{H}'_{fs}$ and $ \hat{H}'_{Z}$ are treated on equal footing (i.e., energy corrections due to the two terms are comparable ${E}'_{fs} \approx {E}'_{Z}$), neither a basis consisting of states in the coupled representation nor a basis consisting of states in the uncoupled representation forms a {\it good} angular basis to find perturbative corrections for the hydrogen atom placed in an external magnetic field. For example, in a basis consisting of states in the coupled representation ($| l, \ j, \ m_j\rangle$), the perturbation matrix $\hat{H}' =\hat{H}'_Z + \hat{H}'_{fs}$ corresponding to the $n=2$ subspace is given below (in which $\gamma = \left( \frac{\alpha}{8}\right)^2$ 13.6 eV, $\alpha = \frac{e^2}{4 \pi \epsilon_0 \hbar c}$, $\beta = \mu_B B_{ext}$ and the basis states are chosen in the order $|0, \ \frac{1}{2}, \ \frac{1}{2}\rangle$, $|0, \ \frac{1}{2}, \ -\frac{1}{2}\rangle$, $|1, \ \frac{3}{2}, \ \frac{3}{2}\rangle $, $|1, \ \frac{3}{2}, \ -\frac{3}{2}\rangle $, $|1, \ \frac{3}{2}, \ \frac{1}{2}\rangle $, $|1, \ \frac{1}{2}, \ \frac{1}{2}\rangle$, $|1, \ \frac{3}{2}, \ -\frac{1}{2}\rangle $, and $|1, \ \frac{1}{2}, \ -\frac{1}{2}\rangle$):
\resizebox{\columnwidth}{!}{
$\hat{H'}=
\left[
\begin{array}{cccccccc}
5\gamma - \beta &0&0&0&0&0&0&0\\
0&5\gamma + \beta&0&0&0&0&0&0\\
0&0&\gamma-2\beta&0&0&0&0&0\\
0&0&0&\gamma+2\beta&0&0&0&0\\
0&0&0&0&\gamma -\frac{2}{3}\beta&\frac{\sqrt{2}}{3}\beta&0&0\\
0&0&0&0&\frac{\sqrt{2}}{3}\beta&5\gamma -\frac{1}{3}\beta&0&0\\
0&0&0&0&0&0&\gamma+\frac{2}{3}\beta&\frac{\sqrt{2}}{3}\beta\\
0&0&0&0&0&0&\frac{\sqrt{2}}{3}\beta&5\gamma+\frac{1}{3}\beta\\
\end{array}
\right]$.}

The following procedure describes what students should be able to do to determine a {\it good} basis and find the first order corrections to the energy spectrum for the Zeeman effect: (1) choose an initial basis consisting of a complete set of eigenstates of $\hat{H}^0$, %(e.g., one is free to choose an angular basis consisting of states in the coupled representation or a basis consisting of states in the uncoupled representation or any other basis for a fixed $n$), 
(2) write the $\hat{H}^0$ and $\hat{H}'$ matrices in the initially chosen basis, (3) recognize $\hat{H}'$ in each degenerate subspace of $\hat{H}^0$,  
%, then if $\hat{H}'$ is diagonal in each degenerate subspace of $\hat{H}^0$ then the orginally chosen basis is a {\it good} basis, otherwise the initially chosen basis is not a {\it good} basis (4) if the initially chosen basis is not a {\it good} basis, 
(4) diagonalize the $\hat{H}'$ matrix in each degenerate subspace of $\hat{H}^0$ to determine a {\it good} basis, and (5) identify and be able to explain why the first-order corrections to the energy spectrum are the diagonal matrix elements of the $\hat{H}'$ matrix given by $E_n^1 = \langle \psi_n^0|\hat{H}'|\psi_n^0\rangle$ in the {\it good} basis. 

\vspace*{-.05in}

\section{Methodology}% for Investigating Student Difficulties}
\vspace*{-.05in}

Student difficulties with the corrections to the energies of the hydrogen atom for the Zeeman effect using DPT were investigated using five years of data involving responses from 64 upper-level undergraduate students and 42 first-year graduate students to open-ended and multiple-choice questions administered {in-class} after traditional instruction in relevant concepts. The undergraduates were in an upper-level QM course, and graduate students were in a graduate-level QM course. Additional insight about the difficulties was gained from 13 individual think-aloud interviews (a total of 45 hours) {with undergraduate and graduate students following the completion of their quantum mechanics courses}. 
%Only at the end were they asked to clarify their responses. 
Students were provided with all relevant information discussed in the introduction section and had lecture-based instruction in relevant concepts. Similar percentages of undergraduate and graduate students displayed difficulties with DPT.

We first analyzed responses of 32 undergraduates on questions related to DPT in the context of the Zeeman effect for the hydrogen atom administered in two previous years. 
Then, we examined the difficulties that 32 undergraduate and 42 graduate students had with identifying a {\it good} basis for the Zeeman effect in the following three years as part of an in-class quiz after traditional lecture-based instruction.  %In all cases, students were told that the radial part of the basis is $R_{nl}(r)$.  
The following question is representative of a series of questions that were posed after traditional lecture-based instruction on relevant concepts (the operator $\hat{H}',$ in Q1, is a proxy for the operators
% $\hat{H}'_r$, $\hat{H}'_{SO}$, 
$\hat{H}'_{fs}$, $\hat{H}'_Z$, and $\hat{H}'_{fs}+\hat{H}'_Z$ that were listed individually in three separate questions):\\
\noindent{\it {\bf Q1.} A perturbation $\hat{H}'$ acts on a hydrogen atom with the unperturbed Hamiltonian $\hat{H}^0 = -\frac{\hbar^2}{2m} \nabla^2 - \frac{e^2}{4 \pi \epsilon_0}\left(\frac{1}{r}\right)$. For the perturbation Hamiltonian $\hat{H}'$, circle \underline{\bf ALL} of the representations that form the angular part of a {\it good} basis and explain your reasoning. Assume that for all cases the principal quantum number is fixed to $n=2$.\\
%($C$ is a constant which makes the dimensions of $\hat{H}'$ that of energy in each case.) \\
%\item A perturbation $\hat{H}'$ acts on a hydrogen atom with the unperturbed Hamiltonian $\hat{H}^0 = -\frac{\hbar^2}{2m} \nabla^2 - \frac{e^2}{4 \pi \epsilon_0}\frac{1}{r}$. For each of the following perturbations, state whether the {\it good} basis to find the first-order corrections to the energies is the coupled or the uncoupled representation, or in any orthonormal basis found with linear combinations of a complete set of the coupled or uncoupled states. {\bf You must explain your answer.}\\
%\begin{enumerate}
%\item $\hat{H}^0 = -\frac{\hbar^2}{2m} \nabla^2 - \frac{e^2}{4\pi \epsilon_0}\left(\frac{1}{r}\right)$
%\begin{enumerate}
%\item Coupled representation
%\item Uncoupled representation
%\item Both coupled and uncoupled representation
%\item Neither coupled nor uncoupled representation
%\end{enumerate}
%\item $\hat{H}' = C\delta(r)$ \label{1}
%\begin{enumerate}
i. Coupled representation,\\
ii. Uncoupled representation,\\
%\item Both coupled and uncoupled representation
%\item Neither coupled nor uncoupled representation
iii. Any arbitrary complete orthonormal basis constructed with linear combinations of states in the coupled representation with the same $l$ (i.e., states with different $l$ values are not mixed),\\
iv. Any arbitrary complete orthonormal basis constructed with linear combinations of states in the uncoupled representation with the same $l$ (i.e., states with different $l$ values are not mixed),\\
v. Neither coupled nor uncoupled representation.}
%\end{enumerate}
%\item Explain your answer.\vfill

In order to find the perturbative corrections, one must first choose a {\it good} basis. Q1 focuses on the bases that form a {\it good} angular basis for the perturbation %for the Zeeman effect with 
$\hat{H}'=\hat{H}'_{fs}+\hat{H}'_Z$, as well as the perturbations 
%$\hat{H}'_r$, $\hat{H}'_{SO}$, 
$\hat{H}'_{fs}$ and $\hat{H}'_Z$ individually. Knowledge of the bases that form a {\it good} angular basis for the individual perturbations   
%$\hat{H}'_r$, $\hat{H}'_{SO}$, 
$\hat{H}'_{fs}$ and $\hat{H}'_Z$ can be helpful when determining a {\it good} basis for $\hat{H}'=\hat{H}'_{fs}+\hat{H}'_Z$. 

The unperturbed Hamiltonian $\hat{H}^0$ is spherically symmetric with unperturbed energies only dependent on $n$ and therefore options i, ii, iii, and iv in Q1 all form a complete set of angular eigenstates of $\hat{H}^0$. Therefore, one must consider which set of basis states in Q1 also diagonalize the given perturbation $\hat{H}'$ in each degenerate subspace of the unperturbed Hamiltonian $\hat{H}^0$. %Since the given degenerate subspace of $\hat{H}^0$ corresponds to $n=2$, a {\it good} basis is one in which the perturbation matrix is also diagonal in that subspace. Considering each term of the perturbation separately, 
In each degenerate subspace of $\hat{H}^0$, the fine structure term $\hat{H}'_{fs}$ is diagonal if the basis is chosen to consist of  states in the coupled representation (option i in Q1) and the Zeeman term is diagonal if the basis is chosen to consist of states in the uncoupled representation (option ii in Q1), but not vice versa. Therefore, for the intermediate field Zeeman effect with $\hat{H}'=\hat{H}'_{fs}+\hat{H}'_Z$, neither a basis consisting of states in the coupled representation nor a basis consisting of states in the uncoupled representation forms a {\it good} basis and 
%neither options i or ii form a {\it good} basis. As a basis consisting of states in the coupled representation is not a {\it good} basis, then {\bf ANY} arbitrary orthonormal basis constructed with a linear combination of states in the coupled representation (option iii) cannot be a {\it good} basis. A similar argument can be made for why option iv is not a {\it good} basis. 
option v in Q1 is correct. In order to determine a {\it good} basis, one may first choose a basis, e.g., consisting of states in either the coupled or uncoupled representation and then diagonalize the perturbation $\hat{H}'=\hat{H}'_{fs}+\hat{H}'_Z$ in the $n=2$ degenerate subspace of the unperturbed Hamiltonian $\hat{H}^0$. %Thus, they must first express either the perturbation $\hat{H}'_{fs}$ or $\hat{H}'_Z$ in an initial basis in which it is not diagonal in the degenerate subspace of $\hat{H}^0$. Then, students must be able to diagonalize the perturbation $\hat{H}'=\hat{H}'_{fs}+\hat{H}'_Z$ in the degenerate $n=2$ subspace of $\hat{H}^0$ and be able to find the corrections to the energy spectrum.

\vspace*{-.03in}

\section{Student Difficulties}
\vspace*{-.03in}

Students had some difficulties with DPT in general (not restricted to the context of the Zeeman effect only). For example, when students were asked to determine %the angular part of the wavefunction for 
a {\it good} basis for finding the corrections to the energies of the hydrogen atom, some students did not even realize that DPT should be used. Other students knew that they had to use DPT to find the corrections to the wavefunction, but they did not use DPT to find the first-order corrections to the energies. These students often incorrectly claimed that they did not need to use DPT since no terms in $E_n^1 = \langle \psi_n^0|\hat{H}'|\psi_n^0\rangle$ ``blow up''.  
%In the context of the intermediate field Zeeman effect, s
Other students only focused on the Zeeman term $\hat{H}'_Z$ when asked to determine %the angular part of the wavefunction for 
a {\it good} basis for finding the corrections to the energies. In particular, they did not take into account the fine structure term $\hat{H}'_{fs}$.  %If the fine structure term $\hat{H}'_{fs}$ is neglected, then one can determine the exact energies for $\hat{H}^0+\hat{H}'_Z$ and there is no need for perturbation theory since $[\hat{H}^0, \hat{H}'_Z]=0$.  
However, the fine structure term must be considered when determining the corrections to the unperturbed energy spectrum for the Zeeman effect. %(even in the case of a strong external magnetic field). 
In response to Q1, students struggled to realize that neither a basis consisting of states in the coupled representation nor a basis consisting of states in the uncoupled representation forms a {\it good} basis for the perturbative corrections to the hydrogen atom placed in an external magnetic field. The results are summarized in Table \ref{goodrep}. Table \ref{goodrep} shows that only 44\% of undergraduates and 33\% of graduate students correctly identified that option v in Q1 is the correct answer for the Zeeman effect. Additionally, 16\% of undergraduate and 17\% of graduate students did not provide any answer to question Q1 after traditional instruction in relevant concepts.

\begin{table}[t]%[!tbp]
\caption[Difficulties]{The percentages of undergraduate (U) and graduate (G) students who chose the options i-v in Q1 for the perturbation $\hat{H}'=\hat{H}'_{fs}+\hat{H}'_{Z}$ after traditional instruction for undergraduates (U) (number of students $N=32$) and graduate students (G) ($N=42$). }%.}% listed representations as the correct representations to form a {\it good} angular basis for the unperturbed Hamiltonian $\hat{H}^0$ and the perturbation $\hat{H}'=\hat{H}'_{fs}+\hat{H}'_{Z}$ and the percentage of students who left Q1 blank after traditional lecture-based instruction 
\label{goodrep}
\centering
\label{goodrep}
\centering
\begin{tabular}
%{\linewidth}{>{\itshape}l 
{|c|c|c|c|c|c|c|}
\toprule
\hline
&i&ii&iii&iv&v&Blank\\
\hline
 U (\%) &28&22&16&13&44&16\\ 
\hline
G (\%)&29&17&12&12&33&17\\
\hline
\bottomrule
\end{tabular} %
\vspace*{-.1in}
\end{table}

Below, we discuss student difficulties %students have with the Zeeman effect corrections to the energy spectrum of the hydrogen atom in the context of DPT 
in selecting the representation that forms a {\it good} basis in Q1 and finding the corrections to the energy spectrum, based %. In this section, we focus on the qualitative results found 
primarily upon responses during the think aloud interviews. 

{\bf A. Difficulty understanding why diagonalizing the entire $\hat{H}'$ matrix is problematic:}  
Many students did not realize that when the initially chosen basis is not a {\it good} basis and the unperturbed Hamiltonian $\hat{H}^0$ and the perturbing Hamiltonian $\hat{H}'=\hat{H}'_{fs}+\hat{H}'_Z$ do not commute, they must diagonalize the $\hat{H}'=\hat{H}'_{fs}+\hat{H}'_Z$ matrix {\it only} in each degenerate subspace of $\hat{H}^0$. %{For example, students were given the system with Hamiltonian {\bf H4} in Eq.} \ref{examH2} { on their midterm examination and asked to determine the first order corrections to the energies. In the Hamiltonian {\bf H4}, $\hat{H}^0$ and $\hat{H}'$ do not commute.} 
%On the midterm examination, 9 of the 20 students diagonalized the entire $\hat{H}'$ matrix instead of diagonalizing the $\hat{H}'$ matrix only in the degenerate subspace of $\hat{H}^0$. 
When presented with a similar system and asked to determine the perturbative corrections, one interviewed student who attempted to diagonalize the entire $\hat{H}'$ matrix justified his reasoning by incorrectly stating, ``We must find the simultaneous eigenstates of $\hat{H}^0$ and $\hat{H}'$."  Discussions suggest that this student, and others with similar difficulties often did not realize that when $\hat{H}^0$ and $\hat{H}'=\hat{H}'_{fs}+\hat{H}'_Z$ do not commute, we cannot simultaneously diagonalize $\hat{H}^0$ and $\hat{H}'=\hat{H}'_{fs}+\hat{H}'_Z$ since they do not share a complete set of eigenstates. Students struggled with the fact that if $\hat{H}^0$ and $\hat{H}'=\hat{H}'_{fs}+\hat{H}'_Z$ do not commute, diagonalizing $\hat{H}'=\hat{H}'_{fs}+\hat{H}'_Z$ produces a basis in which $\hat{H}^0$ is not diagonal. However, since $\hat{H}^0$ is the dominant term and $\hat{H}'=\hat{H}'_{fs}+\hat{H}'_Z$ provides only small corrections, we must ensure that the basis states used to determine the perturbative corrections  %in Eqs. \ref{energy} and \ref{wave} 
remain eigenstates of $\hat{H}^0$. 

{\bf B. Incorrectly claiming that  BOTH a basis consisting of states in the coupled representation and a basis consisting of states in the uncoupled representation are {\it good} bases:}  % for the intermediate field Zeeman effect:}   
%Correctly identifying a {\it good} basis for the fine structure term $\hat{H}'_{fs}$ and the Zeeman term $\hat{H}'_{Z}$ but incorrectly identifying a {\it good} basis for the perturbation $\hat{H}'=\hat{H}'_{fs}+\hat{H}'_Z$:} 
%Many students had difficulty identifying a {\it good} basis for perturbative corrections for the intermediate field Zeeman effect. %Since in each degenerate subspace of $\hat{H}^0$, $\hat{H}'_{fs}$ is diagonal if a basis consisting of states in the coupled representation is chosen and $\hat{H}'_Z$ is diagonal if a basis consisting of states in the uncoupled representation is chosen, but not vice versa, neither basis forms a {\it good} basis for the Zeeman effect. 
%For example, i
In Q1, many students correctly identified that the {\it good} basis for the fine structure term $\hat{H}'_{fs}$ is a basis consisting of states in the coupled representation (option i) and also correctly identified that the {\it good} basis for the Zeeman term $\hat{H}'_Z$ is a basis consisting of states in the uncoupled representation (option ii in Q1). However, after correctly identifying the {\it good} basis for the two perturbations individually, some students did not realize that neither the coupled nor the uncoupled representation (option v in Q1) forms a {\it good} basis for the Zeeman effect in which the perturbation is $\hat{H}'=\hat{H}'_{fs}+\hat{H}'_Z$. One interviewed student incorrectly claimed that ``the coupled are a {\it good} basis for $\hat{H}'_{fs}$ and uncoupled are a {\it good} basis for $\hat{H}'_{Z}$, so both coupled and uncoupled form a {\it good} basis for $\hat{H}'_{fs}+\hat{H}'_Z$." This student and others with this type of response thought that since a basis consisting of states in the coupled representation (option i in Q1) forms a {\it good} basis for the fine structure term $\hat{H}'_{fs}$ and a basis consisting of states in the uncoupled representation (option ii in Q1) forms a {\it good} basis for the Zeeman term $\hat{H}'_{Z}$, a {\it good} basis for the perturbation consisting of the sum of these two perturbations is either a basis consisting of states in the coupled or uncoupled representation. 
%The final four difficulties are due in large part to students' underlying difficulties with the fundamental concepts of linear algebra and applying these concepts in the context of QM for finding corrections to the energy spectrum of the hydrogen atom for the Zeeman effect.

{\bf C.  Incorrectly claiming that a {\it good} basis does not exist for the Zeeman effect:} 
%In Q1, some students who correctly identified that the {\it good} basis for the fine structure term $\hat{H}'_{fs}$ is a basis consisting of states in the coupled representation and also correctly identified that a {\it good} basis for the Zeeman term $\hat{H}'_Z$ is a basis consisting of states in the uncoupled representation, correctly chose that neither the coupled nor the uncoupled representation forms a {\it good} basis for the perturbation $\hat{H}'=\hat{H}'_{fs}+\hat{H}'_Z$ (option v in Q1) but then used incorrect reasoning to do so. Two common examples are as follows:
%{\bf \textit{Incorrectly claiming that the coupled and uncoupled representations are the only possible bases for the angular part of the wavefunction}:} 
%Some students incorrectly argued that since neither an angular basis consisting of states in the coupled representation nor a basis consisting of states in the uncoupled representation forms a {\it good} basis, a 
Some students argued that {\it good} basis does not exist for the intermediate field Zeeman effect and struggled to realize that the coupled representation or the uncoupled representation are not the only two possibilities for the angular basis. One interviewed student with this type of reasoning had difficulty understanding options iii and iv in Q1, stating: ``I don't know what a linear combination of coupled or uncoupled states is. I thought there were just coupled states or uncoupled states." This student and others with this type of reasoning  did not realize that a {\it good} basis could be constructed from a linear combination of states in the coupled or uncoupled representation.% (or equivalently of states in the uncoupled representation). 

%{\bf \textit{Incorrectly claiming that diagonalizing $\hat{H}'$ in each degenerate subspace of $\hat{H}^0$ makes the $\hat{H}^0$ matrix non-diagonal}:}  
Some students had difficulty realizing that any linear combination of states from the same degenerate subspace of $\hat{H}^0$ are eigenstates of $\hat{H}^0$. For example, one student who correctly identified that neither the coupled nor the uncoupled representation forms a {\it good} basis for the Zeeman effect argued that ``no {\it good} basis exists since we cannot diagonalize a part of the $\hat{H}'$ matrix ($\hat{H}'$ in the degenerate subspace of $\hat{H}^0$) without affecting the $\hat{H}^0$ matrix." This student and others who provided similar incorrect reasoning claimed that by diagonalizing $\hat{H}'$ in the degenerate subspace of $\hat{H}^0$, the $\hat{H}^0$ matrix would no longer be diagonal. However, due to the degeneracy, any linear combination of states from the same degenerate subspace of $\hat{H}^0$ are eigenstates of $\hat{H}^0$. Therefore, diagonalizing $\hat{H}'$ in the degenerate subspace of $\hat{H}^0$ determines the special linear combination that forms a {\it good} basis.
%{\bf Only coupled {\it good} basis for $\hat{H}'_{fs}$ and only uncoupled {\it good} basis for $\hat{H}'_{Z}$, so there is no {\it good} basis for $\hat{H}'_{fs}+\hat{H}'_Z$ (i.e., there are no other options for a basis for the angular part other than the coupled or uncoupled representations):} In Q4, students with this type of difficulty often correctly selected option v. However, these students used incorrect reasoning in choosing their answer. These students correctly identified that a basis consisting of states in the coupled representation forms a {\it good} basis for the fine structure term $\hat{H}'_{fs}$ and also correctly identified that a basis consisting of states in the uncoupled representation forms a {\it good} basis for the Zeeman term $\hat{H}'_Z$. 

{\bf D.  Incorrectly claiming that the choice of the initial basis affects corrections to the energy spectrum:}  
%Not realizing the initial choice of basis does not affect the corrections to the energy spectrum:}
Of the students who correctly identified that a {\it good} basis for the Zeeman effect consists of special linear combinations of states in the coupled or uncoupled representation, % (or, equivalently, special linear combinations of states in the uncoupled representation), 
some  did not realize that the first order corrections to the energy spectrum would be the same regardless of the initial choice of the basis. %Since neither a basis consisting of states in the coupled representation nor a basis consisting of states in the uncoupled representation forms a {\it good} basis, a 
A {\it good} basis cannot easily be identified at the onset. In order to determine a {\it good} basis and the first order corrections to the energy spectrum due to the Zeeman effect, one can initially choose a basis consisting of states in the coupled representation and then diagonalize $\hat{H}'=\hat{H}'_{fs}+\hat{H}'_Z$ in each degenerate subspace of $\hat{H}^0$. However, one could also initially choose a basis consisting of states in the uncoupled representation and then diagonalize $\hat{H}'=\hat{H}'_{fs}+\hat{H}'_Z$ in each degenerate subspace of $\hat{H}^0$ to determine a {\it good} basis and the first order corrections to the energy spectrum due to the Zeeman effect. Regardless of the choice of the initial basis, after diagonalizing $\hat{H}'=\hat{H}'_{fs}+\hat{H}'_Z$ in each degenerate subspace of $\hat{H}^0$, the first order corrections to the energy spectrum due to the Zeeman effect will be the same in any {\it good} basis. Many students thought that the first order corrections to the energies depend on the initial choice of basis. Therefore, if one chooses a basis consisting of states in the coupled representation, then the first order corrections in this case would be different than those obtained had a basis consisting of states in the uncoupled representation been chosen as the initial basis. However, it does not make sense experimentally that the observed perturbative corrections would depend upon the choice of basis. Lack of appropriate connection between physics and mathematics in the context of DPT for the Zeeman effect sheds light on student epistemology and the difficulty in mathematical sense-making in QM \cite{tuminaro}. 

\vspace*{-.12in}
\section{SUMMARY AND FUTURE PLAN}
\vspace*{-.12in}

Both upper-level undergraduate and graduate students struggled with finding perturbative corrections %to the energy spectrum of the hydrogen atom 
to the hydrogen atom energy spectrum for the intermediate field Zeeman effect using DPT. 
 Interviewed students' responses suggested that some of them held epistemological beliefs inconsistent with the framework of QM and struggled with mathematical sense-making in the context of QM in which the paradigm is novel \cite{singh4}. After traditional instruction, some students claimed that different initial choices of the basis before a {\it good}  basis has been found will yield different corrections to the energy spectrum of the hydrogen atom for the Zeeman effect. These students had difficulty in connecting experimental observations with quantum theory and in correctly reasoning that since the corrections to the energy spectrum can be measured experimentally, different choices of the initial basis cannot yield different physically observable corrections to the energy spectrum. Since students are still developing expertise in QM and the DPT requires appropriate integration of mathematical and physical concepts, cognitive overload can be high while reasoning about these problems \cite{sweller}. Many advanced students found it challenging to do metacognition \cite{sweller} in this context of QM and provided responses that were not consistent with each other.
We are using the difficulties as a guide in developing a Quantum Interactive Learning Tutorial (QuILT) to help students develop a {\it good} grasp of these concepts. %After engaging with the QuILT, student interviews and written responses suggest that they were able to reason about DPT more consistently and articulate that one is free to choose either the coupled or uncoupled representation as the initial angular basis and then can proceed to explicitly diagonalize the perturbation $\hat{H}'=\hat{H}'_{fs}+\hat{H}'_Z$ in each degenerate subspace of $\hat{H}^0$ in order to determine a {\it good} basis and the perturbative corrections.  
%We identified several difficulties that students have with these concepts, and we have been using the difficulties found as resources in developing a quantum interactive learning tutorial (QuILT) on DPT in the context of the hydrogen atom. 
%The initial results from the QuILT implementation are encouraging. 

\vspace*{-.09in}
%\section*{ACKNOWLEDGEMENTS}
\begin{acknowledgments}
\vspace*{-.05in}

We thank the NSF for awards PHY-1505460 and 1806691. %We are also thankful to %members of the department of physics and astronomy at the University of Pittsburgh especially 
%R. P. Devaty. %Additionally, we thank the students who interviewed to help improve the QuILT and our understanding of student difficulties.
%\vspace*{-.27in} 
\end{acknowledgments}

\end{document}